\documentclass[12pt,letterpaper]{article}
\pdfoutput=1
\usepackage{opex3}
\usepackage{color}
\usepackage{threeparttable}
\begin{document}

\title{Correlated photon pair generation in AlGaAs nanowaveguides via spontaneous four-wave mixing}

\author{Pisek Kultavewuti,$^{1,\ast}$ Eric Y. Zhu,$^1$ Li Qian,$^1$ Vincenzo Pusino,$^2$ \mbox{Marc Sorel,$^2$} and J. Stewart Aitchison$^1$}

\address{$^1$Department of Electrical and Computer Engineering, University of Toronto, \mbox{10 King's College Road}, Toronto, Ontario  M5S 3G4, Canada\\
$^2$School of Engineering, University of Glasgow, Glasgow  G12 8QQ, Scotland, UK}

\email{$^*$pisek.kultavewuti@mail.utoronto.ca} %% email address is required



\begin{abstract*}
We demonstrate a source of correlated photon pairs which will have applications in future integrated quantum photonic circuits. The source utilizes spontaneous four-wave mixing (SFWM) in a dispersion-engineered nanowaveguide made of AlGaAs, which has merits of negligible two-photon absorption and low spontaneous Raman scattering (SpRS). We observe a coincidence-to-accidental (CAR) ratio up to 177, mainly limited by propagation losses. Experimental results agree well with theoretical predictions of the SFWM photon pair generation and the SpRS noise photon generation. We also study the effects from the SpRS, propagation losses, and waveguide lengths on the quality of our source.
\end{abstract*}

%\ocis{(270.0270) Quantum optics; (190.4390) Nonlinear optics, integrated optics; (190.4380) Nonlinear optics, four-wave mixing.}
%For a complete list of OCIS codes, visit: http://www.opticsinfobase.org/submit/ocis/

%%%%%%%%%%%%%%%%%%%%%%%%%%  body  %%%%%%%%%%%%%%%%%%%%%%%%%%
\linespread{1.3}\selectfont
\section{Introduction}
Correlated photon pairs have potential applications in future communications and information processing systems. Recent demonstrations utilizing entangled photons have been reported  including secure communications \cite{Ursin2007}, teleportation \cite{Bussieres2014}, and quantum computing \cite{Barz2012}. Despite these impressive results, they rely on free-space optics, which hinders scaling and applying the technology beyond the lab. In addition, bulk systems are prone to external interference and are sensitive to perturbations. Recently, there has been a growing interest in the use of guided-wave integrated photonics to realize a compact, robust quantum circuit that contains all necessary components from photon sources to routing elements, interferometers, and photon detectors. The integrated platform offers inherent stability in interfering quantum states, which is the crux of several quantum algorithms.

The III-V semiconductor material AlGaAs/GaAs is a promising candidate for the implementation of complete integrated quantum photonic circuits. The integration of an optical pump source can be achieved significantly easier in this material system than in silicon-based materials \cite{Boitier2014}. An active interferometer circuit \cite{Wang2014a} and a GaAs-based single photon detector \cite{Gaggero2010} have also been demonstrated. Integrated sources of correlated photon pairs in AlGaAs have been rigoursly studied with device structures including superlattice waveguides \cite{Sarrafi2013a,Sarrafi2013b} and Bragg-reflection waveguides \cite{Boitier2014,Horn2012,Orieux2013}. These structures rely on the second-order susceptibility. To the best of our knowledge, there is no study in the performance of the pair production in AlGaAs-based devices via the third-order nonlinearity, regardless of a vast body of the literature on the classical counterpart. Therefore, in this paper, we focus on the pair generation via the third-order nonlinearity in AlGaAs waveguides.

The generation of correlated photon pairs can be based on either spontaneous parametric down-conversion (SPDC) which makes use of the second-order susceptibility $\chi^{(2)}$ or spontaneous four-wave mixing (SFWM) which harnesses the third-order susceptilibity $\chi^{(3)}$. In SPDC, one pump photon is annihilated and two correlated photons are created with efficiency linearly dependent on the pump power. In general, the pump photon and the resulting photon pair are typically spectrally far apart, leading to (post-filtering) high photon pair purity; however, this approach is difficult to phase-match and susceptible to temporal walk-off. The poor phase matching limits the pair generation bandwidth, and the walk-off degrades temporal indistinguishability of pairs (which would degrade pair entanglement properties) generated at different positions along the nonlinear medium. These difficulties can be mitigated by using a shorter medium but with a compromise on the pair production efficiency. Pair generation in AlGaAs has mostly relied upon the SPDC process \cite{Boitier2014,Sarrafi2013a,Horn2012}, and hence faces these challenges.

On the other hand, SFWM provides correlated photon pair generation with easier phase matching requirements. During the SFWM process, two pump photons are annihilated and two correlated photons are created with an efficiency that depends quadratically on the pump power. Because the pump photons and the pair are spectrally close to one another, the phase matching can be maintained over a broad spectral range and results in a wide pump photon acceptance bandwidth. The photons also travel with an approximately similar group velocity that leads to a minimal walk-off problem. Additionally, SFWM could create simultaneous, independent photon pairs into different wavelength channels over a broad spectral range given a fixed pump photon wavelength. This feature is very attractive in quantum channel multiplexing of the future quantum network \cite{Lim2008a,Arahira2013}.

The AlGaAs alloy has several advantageous properties that can exploit benefits from the SFWM process in making an efficient and bright photon pair source. AlGaAs can make high index contrast waveguides and has high $\chi^{(3)}$ coefficients, which in combination results in a large effective nonlinear coefficient $(\gamma\approx500\;\mathrm{m}^{-1}\mathrm{W}^{-1} \cite{Lacava:2014wk})$. Inefficiencies from two-photon absorption (TPA) and free carrier absorption (FCA) are negligible by a proper choice of the alloy compositions \cite{Aitchison1997,Dolgaleva:2011wb}. Its crystalline structure leads to narrow spontaneous Raman scattering (SpRS), which can be filtered easily. In addition, control in AlGaAs waveguide dispersion allows a wide-band pump photon acceptance and a broad correlated photon pair generation bandwidth. These characteristics, in conjunction with on-chip integrateability, makes AlGaAs a very attractive platform to realize a quantum photonic circuit.

In this work, we demonstrated an integrated correlated photon pair source in a dispersion-engineered AlGaAs nanowaveguide via SFWM. We characterize the pair generation rate, the signal-to-noise ratio (i.e. CAR), and the SpRS noise in our device and analyze the measurement results, which agree well with theoretical results. The interplay between SFWM-generated photon pairs, SpRS noise, propagation loss, and waveguide length is also studied.

\section{Theoretical background and device design}
As described briefly in the previous section, SFWM requires the participating photons to satisfy the conservation of energy. For a frequency-degenerate pumping scheme, two pump photons at $\omega_p$ are are annihilated to yield two correlated photons at the signal frequency $\omega_s$ and the idler frequency $\omega_i$, such that
\begin{equation}
	2\omega_p = \omega_s+\omega_i \label{eq:consv_energy}.
\end{equation}

For efficient and broadband SFWM, the dispersion of the nonlinear waveguide must be carefully engineered such that the phase matching condition is satisfied. A phase matching equation for the degenerate-pump co-polarized SFWM is
\begin{equation}
	\Delta\beta = \beta_s+\beta_i-2\beta_p+2\gamma P \label{eq:phasematching1},
\end{equation}
where $\beta_{p,s,i}$ is the propagation constant of the pump, the signal, and the idler modes, $\gamma$ is the nonlinear coefficient, and $P$ is the input pump power. In the low-dispersion regime, Eq. (\ref{eq:phasematching1}) can be further simplified to
\begin{equation}
	\Delta\beta \approx \beta_{\omega_p}^{(2)}(\Delta\omega)^2+2\gamma P \label{eq:phasematching2}
\end{equation}
where $\beta_{\omega_p}^{(2)} = d^2\beta(\omega)/d\omega^2$ is the second-order dispersion coefficient calculated at the pump frequency and $\Delta\omega=\omega_p-\omega_i$ is the angular frequency detuning. It is evident from Eq. (\ref{eq:phasematching2}) that in order to nullify $\Delta\beta$, the dispersion must be anomalous at the pump wavelength. For the C-band operation (1530-1560 nm), this requirement can be met using a deeply-etched AlGaAs structure.

\begin{figure}[tb]
	\centering
	\includegraphics[width=13cm]{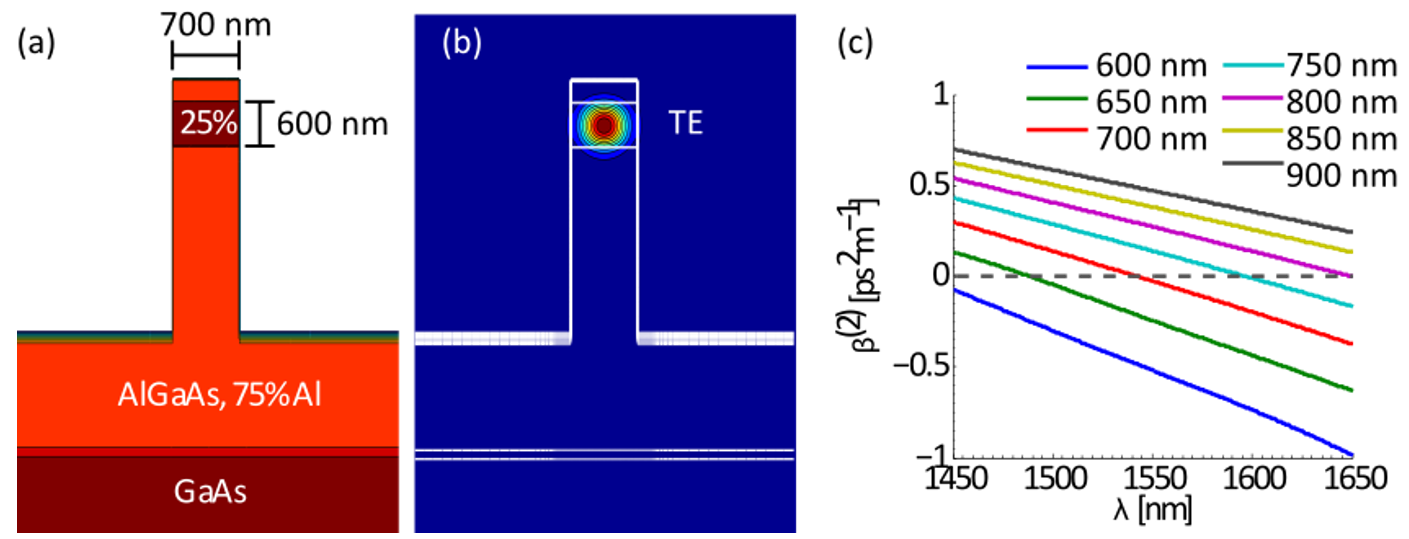}
	\caption{(a) A cross-sectional structure and (b) a simulated TE mode profile of a 700-nm-wide, deeply etched (3000 nm) AlGaAs nanowaveguide. The length of the nanowaveguide is 4.5 mm. (c) The second-order dispersion $\beta^{(2)}$ coefficients of the TE mode in deeply etched AlGaAs waveguides with waveguide widths ranging from 600 nm to 900 nm. All simulations are obtained with Lumerical MODE.}
	\label{fig:cross_section_dispersion}
\end{figure}

Figure \ref{fig:cross_section_dispersion}(a) illustrates the cross-sectional structure of the deeply etched (3000 nm), 700-nm-wide AlGaAs waveguide along with the transverse electric (TE) mode profile in Fig. \ref{fig:cross_section_dispersion}(b). The dispersion of such waveguides can be tuned by varying its width. Figure \ref{fig:cross_section_dispersion}(c) shows the simulated $\beta^{(2)}$ curves of the TE mode for different waveguide widths, and it shows that the 700-nm-wide waveguide becomes anomalous at wavelengths beyond 1550 nm \cite{Meier2007}. For a TE mode, the electric field is oriented horizontally and experiences a high index contrast between AlGaAs and air, which results in tight mode confinement. This effect leads to a significant waveguide dispersion that can outweigh the material dispersion; therefore, anomalous dispersion is achieved.

In the lossless limit, the idler wave that is generated through the SFWM processes has an optical power of \cite{Helt2012}
\begin{equation}
	P_i^\mathrm{SFWM} = \hbar\omega_i B(\gamma PL)^2\,\mathrm{sinc}^2(\Delta\beta L/2) \label{eq:power_i}
\end{equation}
where $B$ is the filter passband (in Hz), $L$ is the waveguide length, and the sinc function is defined as $\mathrm{sinc}=\sin\,x/x$. The (squared) sinc function term in \mbox{Eq. (\ref{eq:power_i})} dictates the efficiency and the bandwidth of the idler generation (or frequency conversion) process, and it is regarded as a phase matching function.

%In Fig. \ref{fig:sinc}, we plot the phase matching function versus signal-pump wavelength detunings, $\lambda_s-\lambda_p$, for both the TE and TM (transverse magnetic) polarizations of the pump wave and pump peak powers of 0, 0.1, and \mbox{0.2 W}, assuming the pump wavelength $\lambda_p=1555\;\mathrm{nm}$, $\gamma=450\;\mathrm{m}^{-1}\mathrm{W}^{-1}$, and $L_\mathrm{eff}=2.8\;\mathrm{mm}$. It is evident that the 700-nm-wide waveguide yields a SFWM  bandwidth of at least 120 nm, for a TE-polarized pump.

%\begin{figure}[tb]
%	\centering
%	\includegraphics[width=8cm]{sinc_detuning_various_powers_v2.pdf}
%	\caption{A plot of the phase matching function (see Eq. (\ref{eq:power_i})) versus wavelength detunings and pump peak powers for the TE (solid curves) and the TM (dash-dot curves) modes of a 700-nm-wide waveguide, given the pump wavelength of 1555 nm, the nonlinear coefficient $\gamma=450\;\mathrm{m}^{-1}\mathrm{W}^{-1}$ and the effective length of $L_\mathrm{eff}=2.8\;\mathrm{mm}$.}
%	\label{fig:sinc}
%\end{figure}

Another benefit of tight mode confinement is a small effective mode area, $A_\mathrm{eff}$, that boosts the nonlinear coefficient according to $\gamma=\beta n_2/A_\mathrm{eff}$, where $n_2$ is the nonlinear refractive index. In effect, the nano-scale waveguides deliver two-fold benefits: broadband phase matching and enhanced nonlinear interaction. However, it is well-known that a large $n_2$ is accompanied by a significant TPA  coefficient $\beta_2$, as both scale with $\chi^{(3)}$. Fortunately, this problem is suppressed in $\mathrm{Al}_x\mathrm{Ga}_{1-x}\mathrm{As}$ as its bandgap is tunable via aluminum concentrations \cite{Aitchison1997}. With $x\sim 0.20$, the bandgap energy corresponds to a photon energy at 720 nm \cite{ElAllali1993}, hence, larger than a two-photon energy at 1550 nm, rendering a two-photon electronic transition \emph{forbidden}. Even though a three-photon transition is still \emph{allowed}, it requires a larger pump power rarely attained in typical SFWM operation. Another benefit of using AlGaAs (which is a direct bandgap material) is that its free carriers have a short lifetime, yielding low FCA and plasma dispersion effects.

\section{Fabrication and classical characterization}
The AlGaAs waveguide structure, whose compositional details can be found in \cite{Kultavewuti2015}, was grown via Metal Organic Vapour Phase Epitaxy (MOVPE) on a GaAs substrate. The nonlinear AlGaAs waveguides were then fabricated using electron-beam lithography patterning of HSQ resist followed by a SiCl$_4$-based reactive ion etching step to an etch depth of 3 micrometers. The resulting nonlinear nanowaveguide was 700-nm wide and 4.5-mm long, matching its designed dimensions. The ends of the nanowaveguide were tapered out to 2-$\mu$m-wide waveguides to facilitate coupling. The propagation loss of the fundamental mode was $\alpha=10\;\mathrm{dB/cm}$. which was measured using the Fabry-Perot technique. Therefore, the propagation loss contribution from the nanowaveguide was 4.5 dB. The total loss from the taper sections was approximately 1 dB.

\begin{figure}[tb]
	\centering
	\includegraphics[width=15cm]{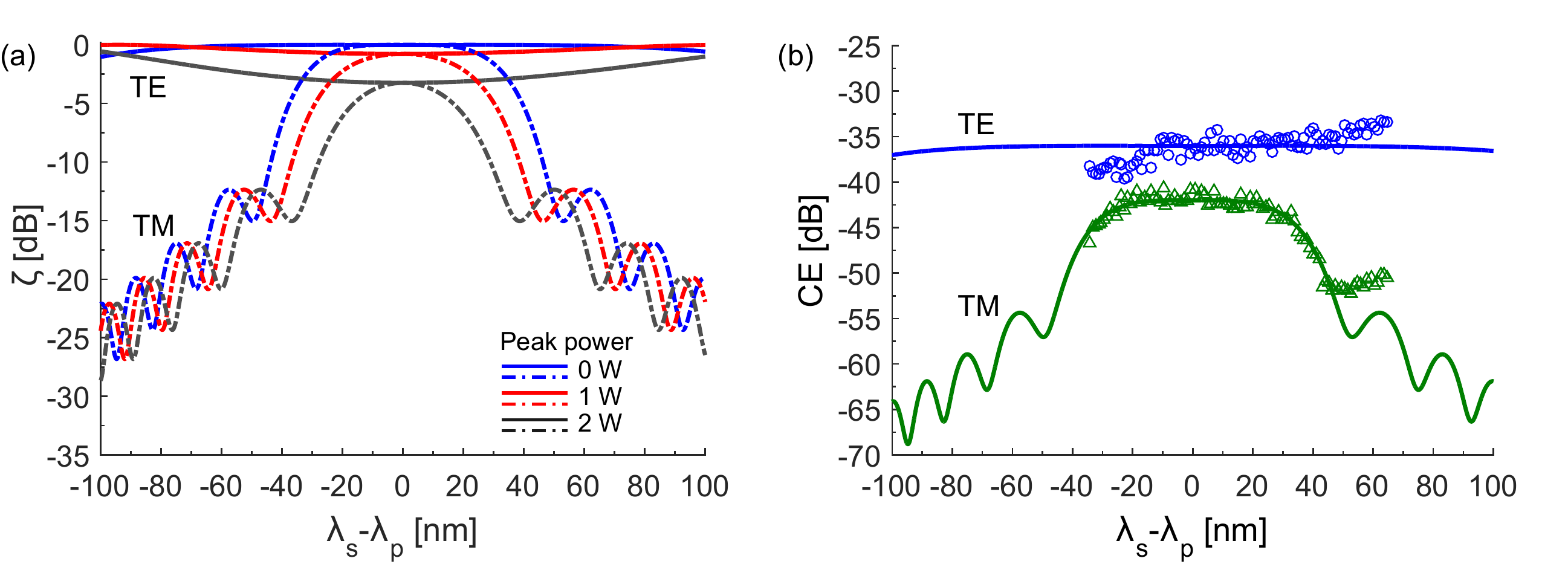}
	\caption{(a) The loss-included phase matching function $\zeta$ (see Eq. (\ref{eq:zeta})) versus wavelength detunings and pump peak powers for the TE (solid curves) and the TM (dash-dot curves) modes of a 700-nm-wide waveguide, given the pump wavelength of 1555 nm, $\gamma=150\;\mathrm{m}^{-1}\mathrm{W}^{-1}$, $\alpha=10\;\mathrm{dB/cm}$, and $L=4.5\;\mathrm{mm}$. (b) Plots of conversion efficiency, CE, from classical cw-FWM measurements and from theoretical predictions for both TE (blue circles for data and curve for theory) and TM (green triangles and curve). Note the discrepancy between the measured data and the theoretical plot for the TM mode is due to our OSA's limited sensitivity.} \label{fig:cw_FWM}
\end{figure}

In order to gauge the possible SFWM bandwidth of our device, we characterize the device using classical pump-degenerate four-wave mixing (FWM), as its bandwidth characteristics are related to that of the SFWM process \cite{Helt2015}. In the classical domain, the generated idler power is expressed as \cite{Shibata1987}
\begin{equation}
	P_{i,\mathrm{out}}^\mathrm{cls} = P_{s,\mathrm{in}}(\gamma PL_\mathrm{eff})^2\,e^{-\alpha L}\zeta \label{eq:piclassical}
\end{equation}
where $P_{s,\mathrm{in}}$ is the input signal power, $L_\mathrm{eff}=[1-\mathrm{exp}(-\alpha L)]/\alpha$ is the nonlinear effective length, and $\zeta$ is regarded as the loss-included phase matching function and is defined by
\begin{equation}
\zeta = \frac{\alpha^2}{\alpha^2+(\Delta\beta)^2}\Bigg(1+\frac{4 e^{-\alpha L}\sin^2(\Delta\beta L/2)}{(1-e^{-\alpha L})^2}\Bigg). \label{eq:zeta}
\end{equation}
The shapes of $\zeta$ for the TE and TM (transverse magnetic) modes are plotted in Fig. \ref{fig:cw_FWM}(a) versus signal-pump wavelength detunings at pump peak powers of 0, 1, and 2 W, assuming the pump wavelength $\lambda_p=1555\;\mathrm{nm}$, $\gamma=150\;\mathrm{m}^{-1}\mathrm{W}^{-1}$, $\alpha=10\;\mathrm{dB/cm}$, and $L=4.5\;\mathrm{mm}$, corresponding to $L_\mathrm{eff}=2.8\;\mathrm{mm}$.

The FWM measurement was performed by introducing a linearly-polarized strong pump (at a fixed wavelength of 1555 nm) and a weak co-polarized  signal, whose wavelength was swept. Both the pump and the signal waves are continuous-wave (cw). The idler wavelength and power were monitored with a calibrated optical spectrum analyzer (OSA). The FWM conversion efficiency is defined as  a ratio of the output idler power to the input signal power $(\mathrm{CE}=P_{i,\mathrm{out}}/P_{s,\mathrm{in}})$. In Fig. \ref{fig:cw_FWM}(b), the measured CEs are plotted for both TE (blue circle) and TM (green triangle) polarization states against their respected calculated CEs. When the pump and the signal waves are in the TE mode, the FWM bandwidth is at least 80 nm, limited by our sources. For the TM mode, the FWM bandwidth is approximately \mbox{60 nm} because the mode is normally dispersive \cite{Kultavewuti2015} over the wavelength of interest. The difference in the maximum CE values of the TE and TM modes is due to different pump powers we could reach with our equipment. The deviation from the theoretical line for the TM polarization is caused by the limited sensitivity of the OSA. From this, we estimate a TE-mode SFWM bandwidth of 80 nm.

In the actual SFWM experiment, we only focus on co-polarized TE-mode SFWM where both the signal and the idler photons are created in the TE modes of the waveguide. Therefore, the difference in the spectral profiles of $\zeta$ of the TE and TM modes is not an issue. However, we would consider the difference in detail in the Discussion section.

\section{SFWM experimentation and results}
\begin{figure}
	\centering
	\includegraphics[width=8cm]{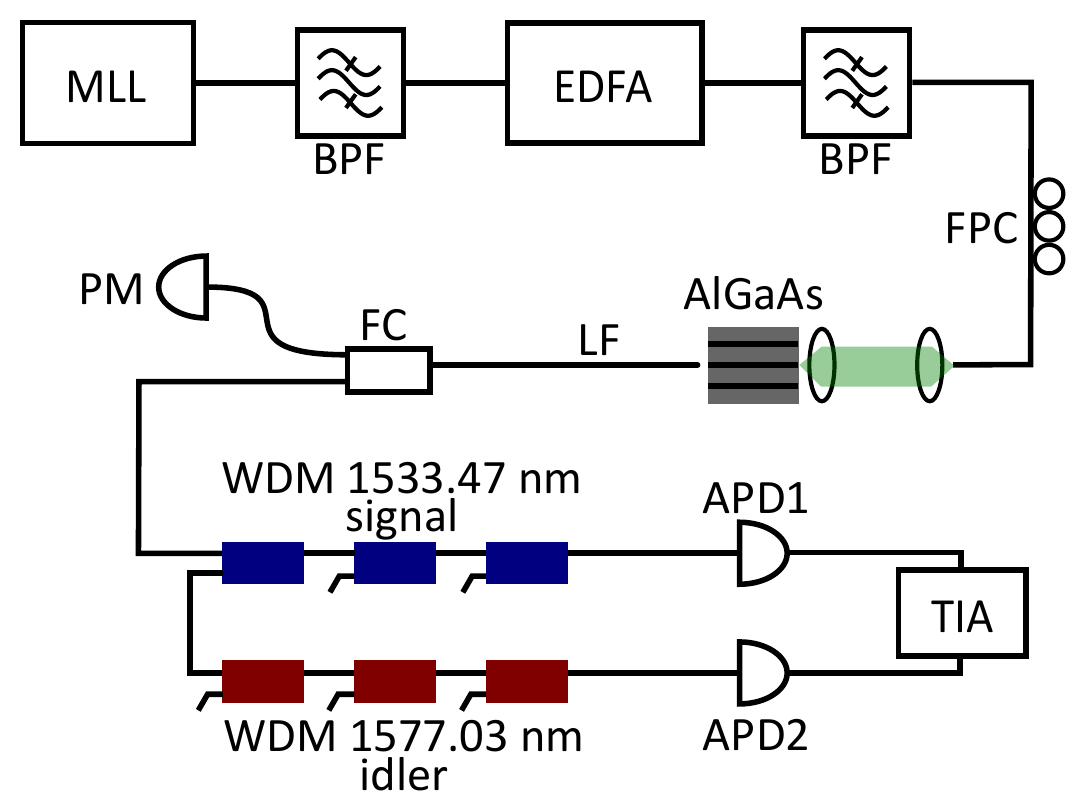}
	\caption{A shematic diagram of the experimental setup for the spontaneous four wave mixing measurement, including: a modelocked laser (MLL), a bandpass filter (BPF), an Erbium-doped fiber amplifier (EDFA), a fiber polarization controller (FPC), a lensed fiber (LF), a fiber coupler (FC), a power meter (PM), an avalanche photodiode (APD), and  a time interval analyzer (TIA).}
	\label{fig:experiment}
\end{figure}

A schematic of the SFWM experiment is shown in Fig. \ref{fig:experiment}. The fiber modelocked laser emits pump pulses with a central wavelength of 1554.9 nm at a rate of 10 MHz. The pump beam passes through a 1-nm bandpass filter to broaden its pulse width to 6 ps, and then an Erbium-doped fiber amplifier (EDFA) to amplify its average power. After this, the beam is transmitted through a pair of cascaded 1-nm bandpass filters to remove any amplified spontaneous emission photons (ASE) from the EDFA. The pump is then coupled into the 4.5-mm-long, 700-nm-wide nanowaveguide via a 40x objective lens. The polarization state of the pump is linear and horizontal to excite only the TE mode in the waveguide. The pump pulses create correlated SFWM photon pairs as they propagate through the waveguide. The output light exits the waveguide and is collected with a lensed fiber (-5.6 dB efficiency). It then enters a 95/5 fiber coupler so that the small portion of the pump light can be monitored with a power meter. The light from the 95$\%$ port passes through a series of fiber-based filters that (1) separate correlated pairs to bands around 1533.47 nm (signal) and 1577.03 nm (idler), each with a 1-nm passband, and (2) filter out the pump photons with a rejection ratio of more than \mbox{110 dB}. The frequency-conjugate signal and idler photons are detected by two InGaAs avalanche photodiodes (APD). Detection clicks are analyzed using an FPGA-based time interval analyzer (TIA) whose temporal resolution is 500 ps (dominated by jitters in the electronics). The two detectors are set to a detection efficiency of 20$\%$ and a deadtime of 15 $\mu\mathrm{s}$ to eliminate counts due to afterpulsing. The coincidence window is $\tau_w=1.5\;\mathrm{ns}$ including two time bins adjacent to the coincidence peak time bin.

We measure the correlated photon pair generation against pump powers. At each pump power, the integration time was set to 90 seconds. Background noise measurement suggests that the SpRS photon noise from fiber components (with nanowaveguides removed from the setup) is negligible in our setup compared to the SpRS noise generated from the AlGaAs waveguide. Additionally, this measurement provides dark count rates of $D_s=1.8\;\mathrm{kHz}$ and $D_i=2.3\;\mathrm{kHz}$.

We also analyze matching of the frequency-conjugate filters by comparing their transmission spectra. In an ideal case, we expect that if a signal photon is transmitted by the signal filter, so should the correlated idler photon be transmitted by the idler filter. In practice, the two filters might not be perfectly frequency-conjugate. The degree of conjugation is defined here as
\begin{equation}
	f = \frac{\int T_sT_i\;d\nu}{\Big(\int T_s\;d\nu\Big)\Big(\int T_i\;d\nu\Big)}, \label{eq:filtermatch}
\end{equation}
where $T_{s,i}$ is the transmission spectrum of the signal or of idler filters. We measured the transmission spectra of our filters and calculated that the conjugation is at $f=0.95$, almost perfect.

Due to the detector deadtime, the measured detection rate must be normalized appropriately by the duty cycle by using the following relation \cite{Knoll1999}
\begin{equation}
	r_\mathrm{actual} = \frac{r_m}{1-\tau_\mathrm{dt}r_m}, \label{eq:ratenormalization}
\end{equation}
where $r_m$ is the measured rate, $r_\mathrm{actual}$ is the actual rate, and $\tau_\mathrm{dt}$ is the detector deadtime. Onward we normalize all measured rates to their corresponding actual rates unless explicitly specified.

In order to analyze our waveguide behavior, we account for both SFWM pairs and Raman generated noise photons. The SFWM pair generation rate, the Raman photon generation rate, and the singles count rates at the signal and the idler bands are expressed as
\begin{equation}
	\mu = aP^2, \label{eq:mu1}
\end{equation}
\begin{equation}
	r_j = b_j P, \label{eq:r1}
\end{equation}
\begin{equation}
	s_j=\eta_j[\mu+r_j], \label{eq:singles1}
\end{equation}
where the subscript $j=s,i$ denotes the signal or the idler, $\mu$ is the number of SFWM photon pairs per pump pulse, $P$ is the coupled input pump peak power, $r_j$ is the number of Raman noise photons per pump pulse, which scales linearly as the pump power \cite{Lin2007}, and $\eta_{s,i}$ is the effective detection efficiency. The coefficients $a$ and $b_j$ determine the efficiency of photon generation through the SFWM and the Raman process respectively. Note that both $\mu$ and $r_j$ are rates at the end of the waveguide before out-coupling. As a result of dark counts and SpRS noise, the registered coincidence rate contains accidental coincidence counts as well. The coincidence rate (per pulse) expected from the measurement, with fiber-based Raman noise safely neglected, can be described as
\begin{equation}
	C = \eta_s\eta_i\Big[(f\mu)+r_s r_i + \mu r_i +r_s \mu + \mu\frac{d_i}{\eta_i} + \frac{d_s}{\eta_s}\mu + r_s\frac{d_i}{\eta_i} + \frac{d_s}{\eta_s}r_i+ \frac{d_s d_i}{\eta_s\eta_i}\Big]. \label{eq:Ctheory}
\end{equation}
In Eq. (\ref{eq:Ctheory}), only the first term accounts for true coincidence. The term $d_j=\tau_wD_j$ is the dark count rate within the coincidence window. The rate $C$ is therefore a $3^\mathrm{rd}$-order polynomial function of the pump power $P$. An accidental coincidence rate due to the adjacent pump pulse is expressed as
\begin{equation}
	A = \Big[\eta_s(\mu+r_s)+d_s\Big]\Big[\eta_i(\mu+r_i)+d_i\Big]. \label{eq:Atheory}
\end{equation}

\begin{figure}[tb]
	\centering
	\includegraphics[width=15cm]{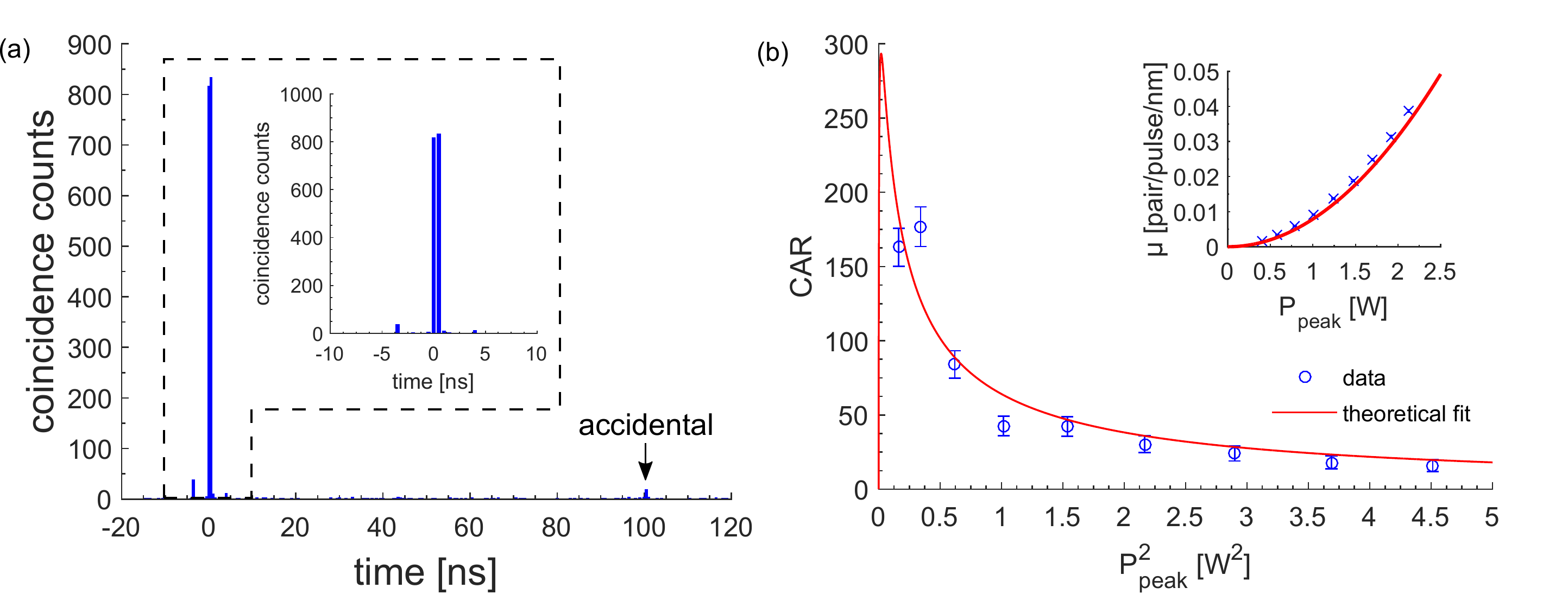}
	\caption{(a) One of the recorded detection coincidence histograms, from which the measured coincidence rate ($C_m$) and the accidental coincidence rate ($A$) were calculated. In particular, it corresponds to the measurement with $\mathrm{CAR}\sim90$, shown in (b). The inset zooms into the counts around the main coincidence peak at 0 ns. (b) Measured and fitted CAR versus coupled input pump peak powers. The inset plots measured and theoretically fitted pair generation rates $\mu$. The integration time is 90 seconds for all measurement points.} \label{fig:his_CAR}
\end{figure}

Experimentally we calculate $C$ and $A$ from the detection coincidence histogram (Fig. \ref{fig:his_CAR}(a)). The main coincidence peak appears at the zero time delay between the signal and the idler. The total measured coincidence rate per second $C_m$ at this peak is determined from the total registered counts within the 1.5-ns coincidence window, divided by the total integration time, and corrected for the detection deadtime. However, $C_m$ also contains the accidental rate $A$. Therefore, $C_m=C+A$. The accidental rate is calculated from counts registered from the next coming pulse, corresponding to a 100-ns delay (not shown in Fig. \ref{fig:his_CAR}(a)). The measured coincidence-to-accidental ratio, CAR, can be defined as
\begin{equation}
	\mathrm{CAR} = \frac{C_m-A}{A}. \label{eq:CARtheory}
\end{equation}

We then seek to extract values for the photon pair generation efficiency $a$, the Raman noise generation rate $b_j$, and the detection efficiencies $\eta_j$ by fitting with the theoretical models in Eq. (\ref{eq:Ctheory})-(\ref{eq:CARtheory}). The optimization fit procedure yields $\eta_s=0.024$, $\eta_i=0.016$, $a=7.6\times10^{-3}\,\mathrm{pair/(pulse}\cdot\mathrm{W}^2)$, $b_s = 2.8\times10^{-3}\,\mathrm{photon/(pulse}\cdot\mathrm{W})$, and $b_i = 2.9\times10^{-3}\,\mathrm{photon/(pulse}\cdot\mathrm{W})$, and the $\mathrm{CAR}$ fit is plotted with the experimental data in Fig. \ref{fig:his_CAR}(b). We observe a CAR of 177 when a pair generation rate is $\mu=29,000\,\mathrm{pair/s/nm}=2.9\times10^{-3}\,\mathrm{pair/pulse/nm}$ at a coupled pump peak power of $0.54\,\mathrm{W}$. The vertical error bar in Fig. \ref{fig:his_CAR}(b) is calculated from the square root of the coincidence counts $\sqrt{C_m}$ by employing the Poissonian nature of the detection. Since the observed rate is $\mu<0.1$, we can safely assume that multipair generation is negligible. Note that at high pump powers, CAR decreases due to the fact that the accidental rate increases. In this regime, the dark count rates $d_j$ are negligible such that CAR varies with $\sim1/\mu$.

\section{Discussion}
We verify that the extracted performance of our waveguide agrees well with theoretical predictions. First, we consider uncorrelated photon noise generated from the SpRS process. Adapted from \cite{Lin2007}, the SpRS photon generation rate over a narrow filtered band is given by
\begin{equation}
	r_\mathrm{theory} = \Bigg[\frac{L_\mathrm{eff}|g_R(\Delta\nu)|N_\mathrm{th}}{A_\mathrm{eff}}\Bigg]B\cdot\tau\cdot P = b_\mathrm{theory} P, \label{eq:ramantheory}
\end{equation}
\begin{equation}
	N_\mathrm{th} = \Bigg\{ \begin{array}{ll} n_\mathrm{th} & \mathrm{;}\;\;\Delta\nu<0, \\ n_\mathrm{th}+1 & \mathrm{;}\;\;\Delta\nu>0, \end{array} \label{eq:phononpopulation1}
\end{equation}
\begin{equation}
	n_\mathrm{th} = \frac{1}{\mathrm{exp}\Big(\frac{h\Delta\nu}{kT}\Big)-1} = 1.85, \label{eq:phononpopulation2}
\end{equation}
where $g_R$ is the Raman gain, $\Delta\nu$ is the frequency detuning from the pump, $\tau=6\;\mathrm{ps}$ is the pump pulse width, and $n_\mathrm{th}$ is the phonon population evaluated at the ambient temperature. In our case, $\Delta\nu=2.7\;\mathrm{THz}$ for the idler band. The filtered bandwidth $B=0.1\;\mathrm{THz}$ corresponds to a 1-nm passband. We found from the experiment that for the idler band $b_i=2.9\times10^{-3}\;\mathrm{photon/(pulse}\cdot\mathrm{W})$ and equating to the factor $b_\mathrm{theory}$ yields $|g_R(\Delta\nu)|=1\times10^{-10}\;\mathrm{cmW}^{-1}$ (equivalent to $2\;\mathrm{m}^{-1}\mathrm{W}^{-1}$), which is in good agreement with Raman gain values reported for AlGaAs \cite{Kao1995}.

For a theoretical prediction of the photon pair generation rate, we adapt Eq. (\ref{eq:power_i}) for pulsed excitation and a lossy waveguide. The effective photon pair rate is expressed as
\begin{equation}
	\mu_\mathrm{theory} = B\tau(\gamma L_\mathrm{eff})^2\;e^{-2\alpha L}\cdot P^2 = a_\mathrm{theory}P^2. \label{eq:mutheory}
\end{equation}
Since \emph{either} of the photon pair members (signal or idler) can be lost due to the linear propagation loss, it is appropriate to capture this effect with the factor $e^{-2\alpha L}$ for the \emph{survived} photon pair rate in Eq. (\ref{eq:mutheory}). The value of $e^{-\alpha L}$ is taken from the linear loss measurement and it includes contributions from the nanowaveguide (4.5 dB; 10 dB/cm, 4.5-mm-long) and the taper sections (1 dB). The effective length is $L_\mathrm{eff}=2.8\;\mathrm{mm}$. The nonlinear coefficient is assumed $\gamma = 150\,\mathrm{m}^{-1}\mathrm{W}^{-1}$, which is consistent with a value observed in a waveguide of a similar structure fabricated using the same process \cite{Kultavewuti2015}. This set of parameters yields $a_\mathrm{theory}=7.5\times10^{-3}\;\mathrm{pair/(pulse}\cdot\mathrm{W}^2)$, and it matches really well with the experimental result.

\begin{figure}[tb]
	\centering
	\includegraphics[width=9cm]{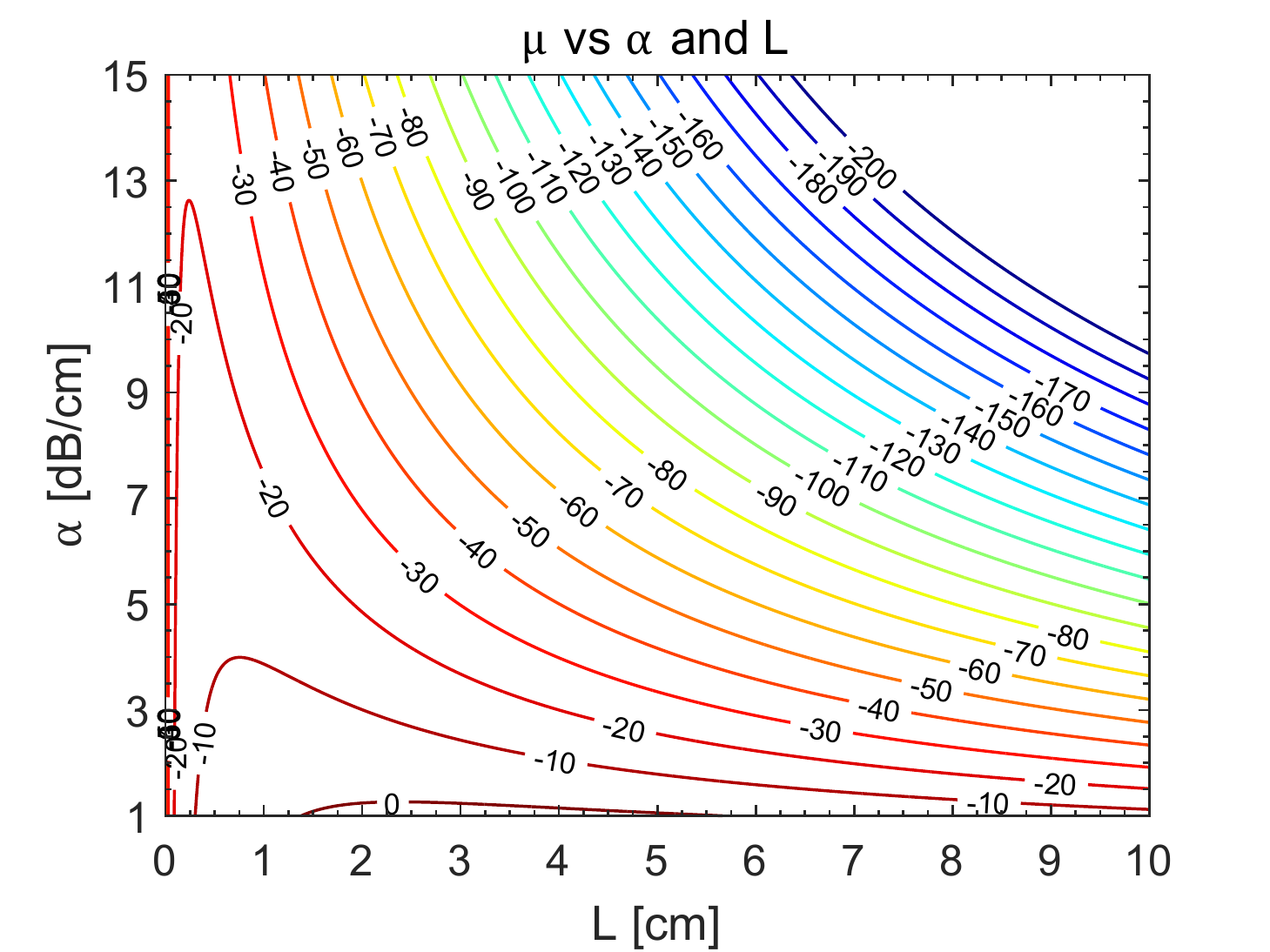}
	\caption{The SFWM pair generation rate $\mu$ (per pump pulse per squared watts, in dB) as a function of propagation losses $\alpha$ (y-axis) and waveguide lengths $L$ (x-axis). Values below -200 dB are omitted.} \label{fig:mu_vs_alpha_L}
\end{figure}

\begin{figure}[tb]
	\centering
	\includegraphics[width=15cm]{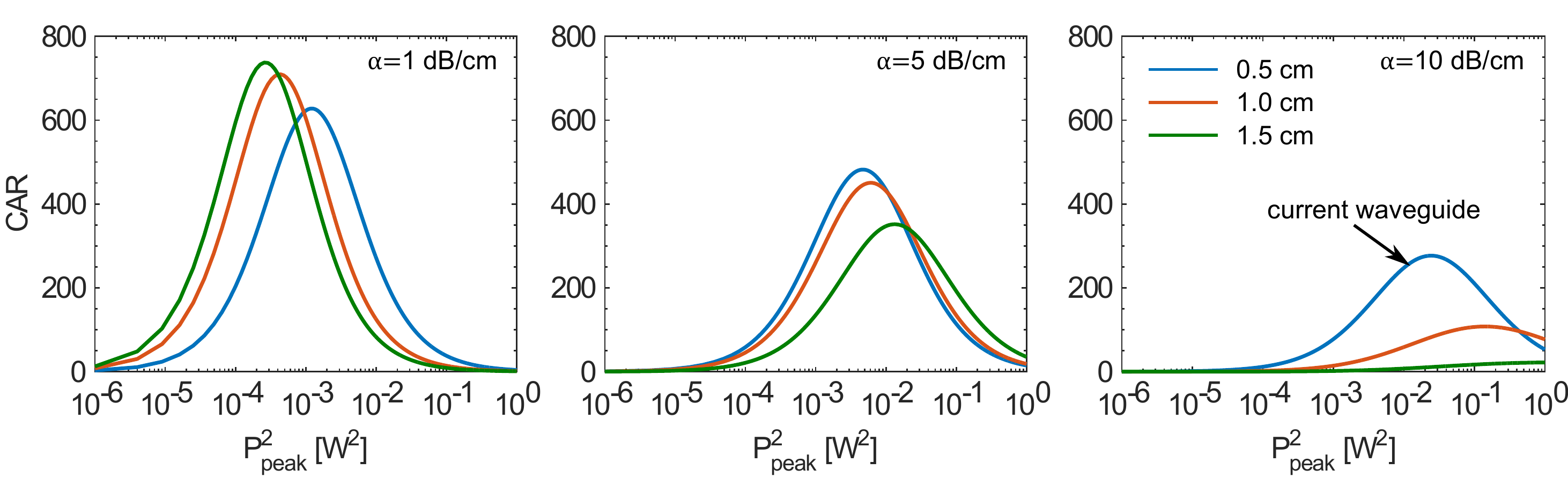}
	\caption{Effects of propagation loss and waveguide length on CAR, calculated using experimentally extracted physical parameters and taking into account the SpRS noise.}
	\label{fig:CAR_effected}
\end{figure}

The performance of our current waveguide is limited by the propagation loss since the pair is lost when one \emph{or} both of the photons are lost. The pair generation rate per squared watts $a_\mathrm{theory}$ is plotted against propagation losses and waveguide lengths in Fig. \ref{fig:mu_vs_alpha_L}. A low propagation loss reduces the power requirement in generating a certain pair rate. It also boosts CAR as the correlated photon pairs are more sensitive (quadratically) to loss than other uncorrelated noise photons. This can be seen in Fig. \ref{fig:CAR_effected} where, using the parameters extracted from the experiment, we plot CAR as a function of pump peak powers, waveguide lengths of 0.5, 1.0, and 1.5 cm, and propagation lossess of 1, 5, and 10 dB/cm. Evidently CAR degrades substantially at high propagation losses. It is worth noting that a propagation loss $\alpha\le 5\,\mathrm{dB/cm}$ is possible for deeply etched AlGaAs nanowaveguides \cite{Apiratikul2014a}.

%We also lost a lot of pairs at the output side where light is coupled from free space to fiber. The coupling loss is about 1.5 dB, which becomes 3 dB for the photon pair. For this aspect, it is possible to improve CAR by replacing the output objective by a lensed fiber to avoid unnecessary free space-to-fiber coupling.

Additionally, in the measurement, we neither observed power-dependent saturation in the single counts nor coincidence counts, which would otherwise indicate effects from TPA and FCA. Usually, in semiconductor-based nonlinear waveguides, TPA and FCA are major causes of saturation effects for parametric processes. However, it has been demonstrated that TPA in AlGaAs materials is very low and no saturation is observed in classical FWM idler generation measurements \cite{Dolgaleva:2011wb,Apiratikul2014a,Pu2010}.

Regarding the difference in phase matching ($\zeta$) profiles of the TE and TM modes as shown in Fig. \ref{fig:cw_FWM}, it would pose a problem to attempts in generating broadband polarization-entangled photon pairs or broadband applications making use of both polarizations. However, as evident from Fig. \ref{fig:cw_FWM}(b), both of the SFWM efficiency curves are quite flat over \mbox{40 nm} and could produce a good polarization-entangled state ($|HH\rangle+e^{i\theta}|VV\rangle$) if both modes are excited at the same pump frequency \cite{Kultavewuti2016}. Therefore, the current waveguide functions well in a single-polarization or narrow-band but dual-polarization operation. It is possible to bring performances of the two modes more similar, for instance, by altering waveguide geometries and compositions. In particular, a square waveguide provides identical dispersion properties for the two modes, and therefore identical conversion efficiency curves.

%\hl{In this paper, we would excite the waveguide with a TE pump and consider only the co-polarized SFWM where both the signal and the idler photons are generated in the TE modes. However, it might be worth considering the differences between the TE and the TM modes. In terms of their spatial profiles, they overlap greatly as long as we operate sufficiently away from the mode cut-off. The most concerning issue is the differential group velocity, which is simulated to be $\sim 0.1$ between the TE and TM modes of wavelengths in the range of 1500 nm to 1600 nm. This might be problematic in applications requiring indistinguishability between photons in the TE and the TM modes. However, in our current situation of the co-polarized SFWM, this fact does not pose an issue. Specifically, the maximum differential group velocity between the pump, the signal, and the idler photons in our case would cause a temporal walk-off of only \mbox{83} fs up on propagation through 5 mm of the waveguide. Nonetheless, there are paths to engineer dispersion properties of the TE and TM modes such that they become more similar. Example would be altering waveguide geometries and compositions.}

SFWM-based correlated photon pair sources have been demonstrated in a number of material platforms including silica \cite{Wang2009,Reimer2015}, chalcogenide \cite{He2012}, crystalline silicon \cite{Harada2010}, and amorphous silicon (a-Si) \cite{Wang2014}. However, silica fibers and chalcogenide glasses suffer from uncorrelated photons originating from the SpRS process, which has a broad gain profile. Crystalline silicon (c-Si), on the other hand, faces with TPA, FCA, and a carrier-induced (plasma dispersion) blue shift \cite{Matsuda2009}. Moreover, these material platforms do not lend themselves to the integration of the pump source.

In Table \ref{tab:compare}, we compare our source with several AlGaAs-based and silicon-based sources. Comparing within the AlGaAs device family, our source, which utilizes $\chi^{(3)}$ rather than $\chi^{(2)}$, has outperformed in the CAR measure and provides a competitive photon pair rate. Considering $\chi^{(3)}$-based devices, our source appears decent in generating photon pair with a high signal-to-noise ratio. Even though the performance we demonstrated here has not matched that of the best-reported c-Si source yet, there is a plenty of room for improvement including lower propagation loss and integrating pump sources. Our study proves the possibility of efficiently generating correlated photon pairs with the AlGaAs material system employing the third-order nonlinearity.

\begin{table}[tb]
	\centering
	\caption{Comparison with other AlGaAs sources and silicon-based sources.}
	\label{tab:compare}
%\begin{center}
{\linespread{1}\selectfont \small
\begin{tabular}{llllrlr}
	\hline
	Work & Material & Structure & Process & CAR & $\mu^a$ & $\nu_s-\nu_i^b$ \\
	\hline
	This work & AlGaAs & waveguide (wg.) & $\chi^{(3)}$ & 177 & $2.9\times10^{-3}$ & 5.4 \\
	Ref. \cite{Sarrafi2013a}$^c$ & AlGaAs & superlattice wg. & $\chi^{(2)}$: Type I & 113 & $2.6\times10^{5}$$^c$ & 5.0 \\
	Ref. \cite{Horn2012} & AlGaAs & Bragg wg. & $\chi^{(2)}$: Type I & 2 & $3.4\times10^{-1}$ & $<0.4$ \\
	Ref. \cite{Orieux2013} & AlGaAs & Bragg wg. & $\chi^{(2)}$: Type II & 19 & $5.8\times10^{-3}$ & $\sim0$ \\
	Ref. \cite{Harada2010} & c-Si & waveguide & $\chi^{(3)}$ & $\sim300$ & $6.5\times10^{-3}$ & 2.8 \\
	Ref. \cite{Wang2014} & a-Si & waveguide & $\chi^{(3)}$ & 170 & $3.4\times10^{-3}$ & 5.4 \\
	\hline
\end{tabular}

$^a$Photon pair generation rate in pair/pulse/nm unless specified otherwise.

$^b$Signal-idler frequency detuning in THz. The value $\sim0$ indicates operation near degeneracy.

\vskip-0.06in$^c$Ref. \cite{Sarrafi2013a} employed a continuous-wave pump. The reported $\mu$ is in pair/s/nm.}
\linespread{1.3}\selectfont
\end{table}

\section{Conclusion}
In summary, we have reported broadband ($>$80 nm) correlated photon pair generation in a dispersion-engineered AlGaAs nanowaveguide utilizing spontaneous four-wave mixing. The CAR of up to 177 is measured with a corresponding pair generation rate of $2.9\times10^{-3}$ pair/pulse/nm. The theoretical models incorporating both SFWM-generated photon pairs and SpRS noise agree well with the experimental results, revealing a feature of low SpRS noise (Raman gain of $g_R=2\;\mathrm{m}^{-1}\mathrm{W}^{-1}$). The nanowaveguide also showed no detrimental effects from two-photon absorption. We project that with lower propagation loss in future waveguides, it is possible to realize brighter correlated photon pair sources with \mbox{CAR$>$300}. This work shows that dispersion-engineered AlGaAs nanowaveguides can serve as a bright correlated photon pair source and offers promises for implementing scalable, integrated quantum photonic circuits.

\section*{Acknowledgment}
We would like to thank the Natural Sciences and Engineering Research Council of Canada
(NSERC) and the Edward S. Rogers Sr. Graduate Scholarship for funding.

%%%%%%%%%%%%%%%%%%%%%%% References %%%%%%%%%%%%%%%%%%%%%%%%%

%\bibliography{bib}{99}

\begin{thebibliography}{10}
\bibitem{Kultavewuti2016}
	We are analyzing and preparing a manuscript on polarization-entangled photon pair generation with orthogonal pumping using the current waveguide design.
\end{thebibliography}
%\bibliographystyle{osajnl}

\end{document}